\documentclass[twocolumn]{webofc}
\usepackage[varg]{txfonts}   % Web of Conferences font
\usepackage{graphicx}  % include figure files
\usepackage{float}
\usepackage{epstopdf}
\usepackage{amsmath}
\usepackage{amssymb}  % for the use of \blacksquare
\usepackage{wasysym}  % for the use of \CIRCLE
\usepackage{xcolor,xspace}
\usepackage{braket}
\usepackage[percent]{overpic}  % insert LaTex text over 
\newcommand{\ba}{\begin{eqnarray}}
\newcommand{\ea}{\end{eqnarray}}
\newcommand{\bsub}{\begin{subequations}}
\newcommand{\esub}{\end{subequations}}

\begin{document}
%
%\title{Insert your title here}
\title{Intertwined quantum phase transitions in odd-mass Nb isotopes}

\author{\firstname{A.} \lastname{Leviatan}\inst{1}\fnsep
  \thanks{\email{ami@phys.huji.ac.il}} \and
  \firstname{N.} \lastname{Gavrielov}\inst{2,3}\fnsep
  \thanks{\email{noam.gavrielov@ganil.fr}}
  \and
\firstname{F.} \lastname{Iachello}\inst{3}\fnsep
\thanks{\email{francesco.iachello@yale.edu}}
}

\institute{Racah Institute of Physics, The Hebrew University, 
Jerusalem 91904, Israel
\and
Grand Acc\'el\'erateur National d'Ions Lourds, CEA/DRF-CNRS/IN2P3,
Bvd Henri Becquerel, BP 55027, F-14076 Caen, France
\and
Center for Theoretical Physics, Sloane Physics 
Laboratory, Yale University, New Haven, Connecticut 
06520-8120, USA
          }

\abstract{%
  A detailed analysis of odd-mass Nb isotopes, 
  in the framework of the interacting boson-fermion model
  with configuration mixing, 
  discloses the effects of an abrupt crossing of states in
  normal and intruder configurations (Type~II QPT),
  on top of  which superimposed
  a gradual evolution from spherical- to 
  deformed-core shapes within the intruder configuration 
  (Type~I QPT). The pronounced presence of both types of 
  QPTs demonstrates, for the first time, the occurrence of
  intertwined QPTs in odd-mass nuclei.
}
\maketitle
\section{Introduction}
\label{intro}
Structural changes induced by variation of parameters
in the Hamiltonian, called quantum phase transitions 
(QPTs), are currently a topic of great interest in
nuclear physics. In this field, most of the
attention has been devoted to the evolution
of structure with nucleon number, exhibiting two types of 
phase transitions. The first, denoted as 
\mbox{Type~I}~\cite{Dieperink1980}, is a shape-phase
transition within a single configuration, as encountered
in the neutron number~90 region~\cite{Cejnar2010}.
The second, denoted as \mbox{Type~II}~\cite{Frank2006},
is a phase transition involving a crossing of different 
configurations, as encountered
in nuclei near (sub-) shell closure~\cite{Heyde11}.
In most cases, the strong mixing between the
configurations obscures the individual QPTs.
However, if the mixing is small,
the Type~II QPT can be accompanied by a distinguished
Type~I QPT within each configuration separately.
Such a scenario, referred to as intertwined QPTs,
was recently shown to occur in the even-even Zr~(Z=40)
isotopes~\cite{Gavrielov2019,Gavrielov2020,Gavrielov2022}.
In the present contribution, we show that a similar
scenario occurs in the adjacent odd-even Nb~(Z=41)
isotopes~\cite{gavleviac22,Gav23}.
\begin{figure}[b]
\centering
\includegraphics[width=\linewidth]{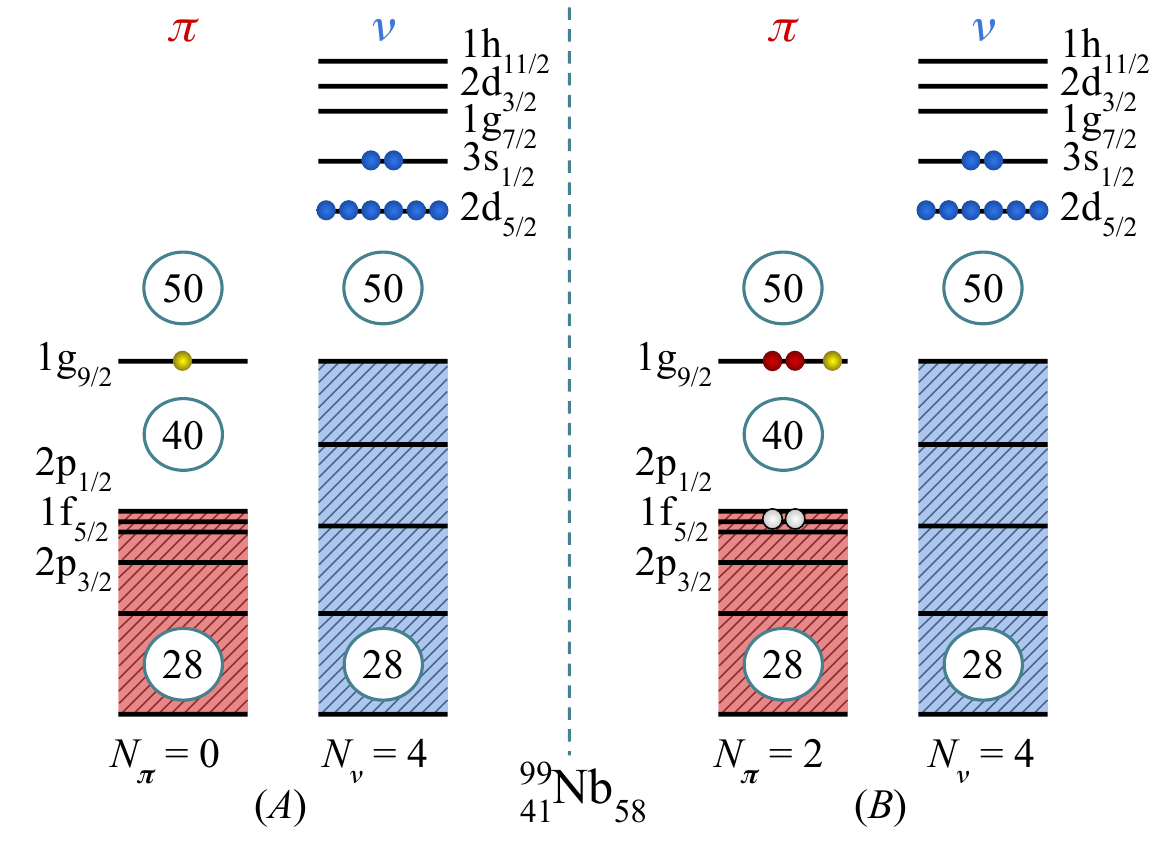}
\caption{Schematic representation of the two coexisting
  shell-model configurations ($A$ and $B$) for
  $^{99}_{41}$Nb$_{58}$. The corresponding numbers of
  proton bosons ($N_{\pi}$) and neutron bosons ($N_{\nu}$),
  are listed for each configuration and $N=N_{\pi}+N_{\nu}$.
  }
\end{figure}
\begin{figure*}[t!]
\centering
\begin{overpic}[width=\linewidth]{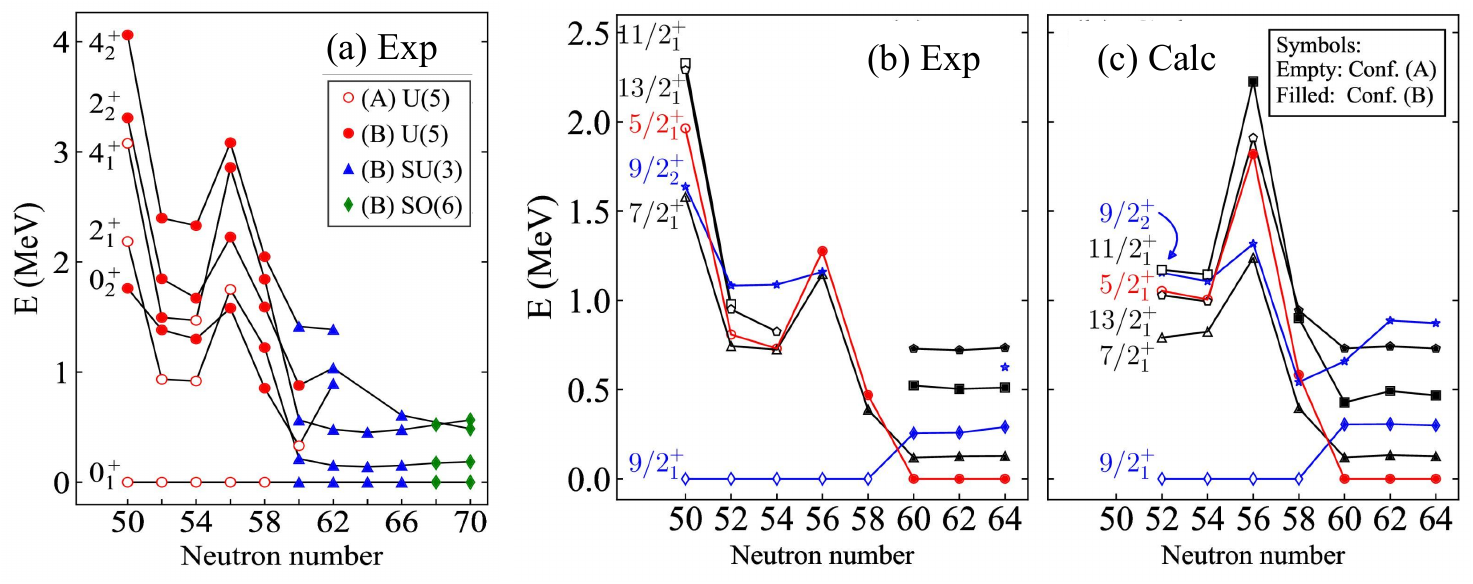} %
\put (17,41) {\large {\bf Zr}}
\put (52,41) { {\large \bf Nb}}
\put (82,41) { {\large \bf Nb}}
\end{overpic}
\caption{Comparison between experimental and calculated
  lowest-energy positive-parity levels in Zr [panel (a)]
  and Nb isotopes [panels (b) and (c)]. Empty (filled) 
symbols indicate a state dominated by the normal 
A~configuration (intruder B~configuration).
In particular, the $0^+_1$ state for Zr and the 
$9/2^+_1$ state for Nb, are assigned to the A (B)
configuration for  
neutron number 52--58 (at and beyond 60). In panel~(a), 
the shape of the symbol [$\circ,\triangle,\diamondsuit$],
indicates the closest dynamical
symmetry [U(5), SU(3), SO(6)] to the level considered.
Note that 
the calculated values start at 52, while the experimental 
values include the closed shell at 50. 
\label{fig:energies-p}}
\end{figure*}
\begin{figure*}[t]
\centering
\begin{overpic}[width=0.49\linewidth]{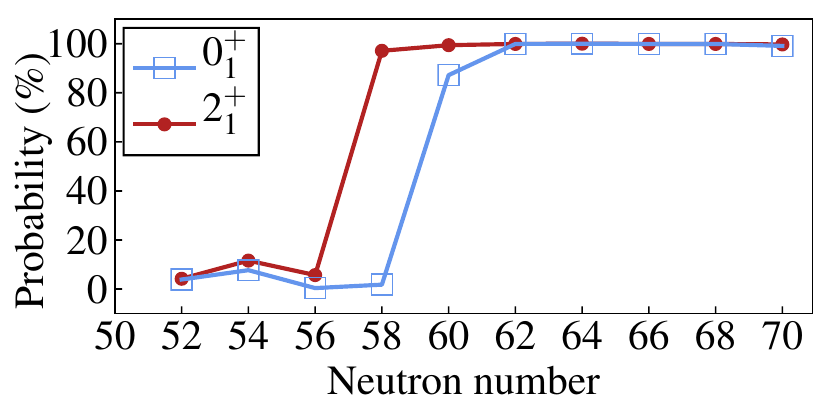}
\put (52,60) {\Large {\bf Zr}}
\put (90,38) {(a)}
\end{overpic}
\begin{overpic}[width=0.49\linewidth]{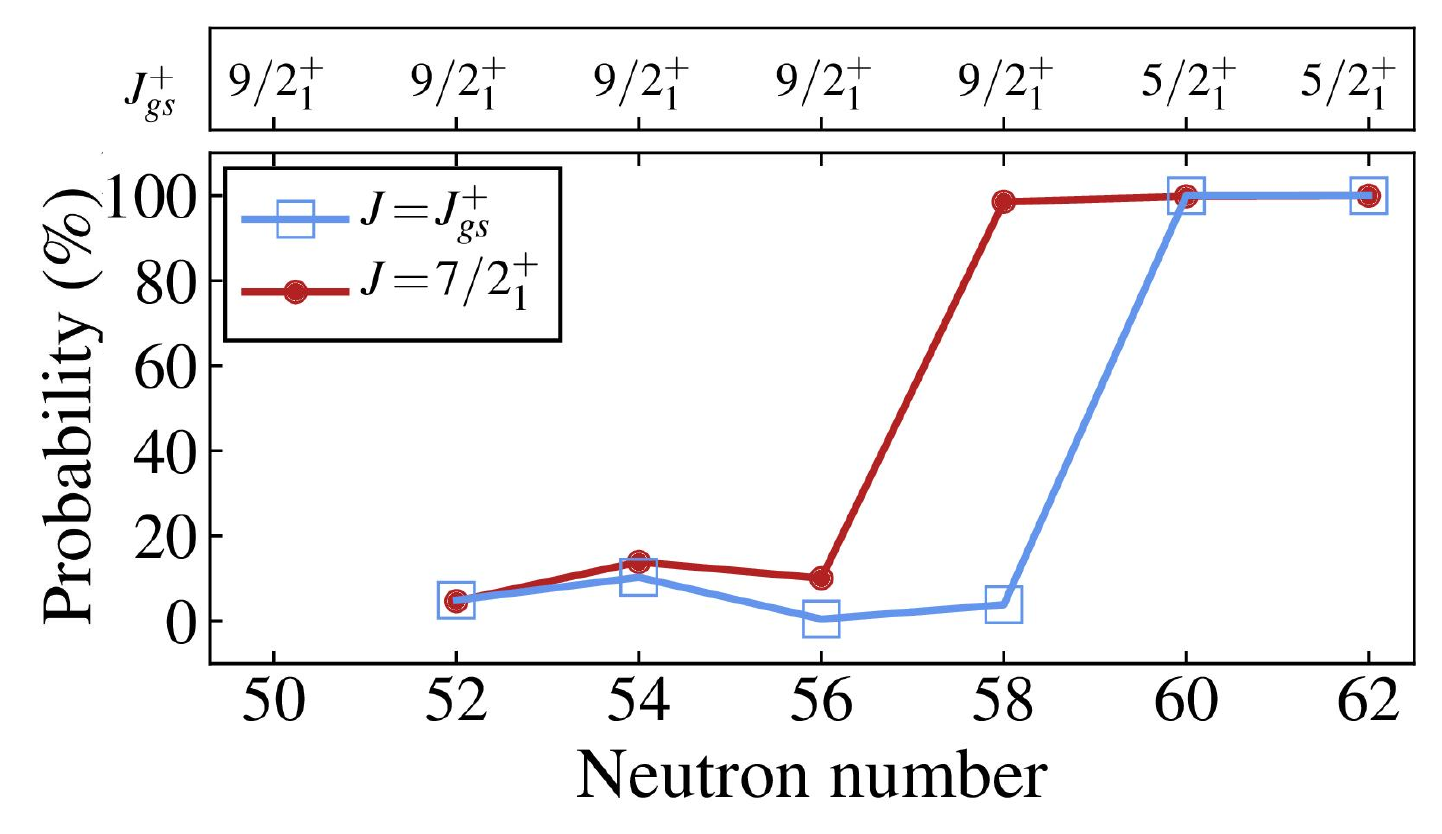}
\put (52,59) {\Large {\bf Nb}}
\put (90,38) {(d)}
\end{overpic}
\begin{overpic}[width=0.49\linewidth]{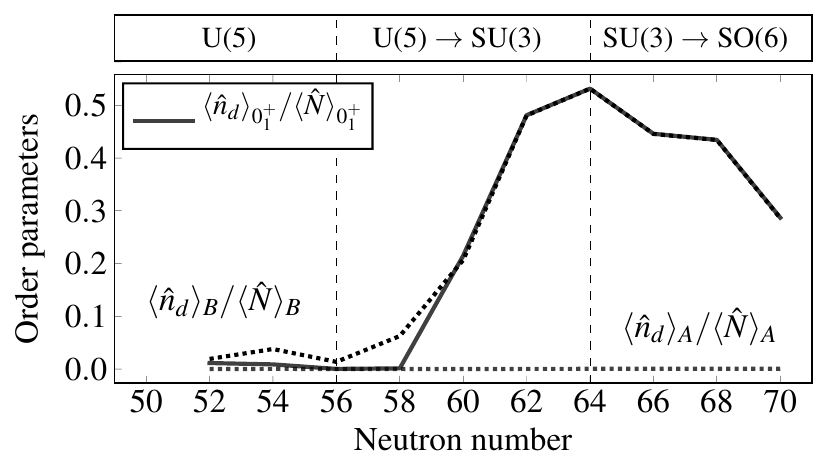}
\put (90,40) {(b)}
\end{overpic}
\begin{overpic}[width=0.49\linewidth]{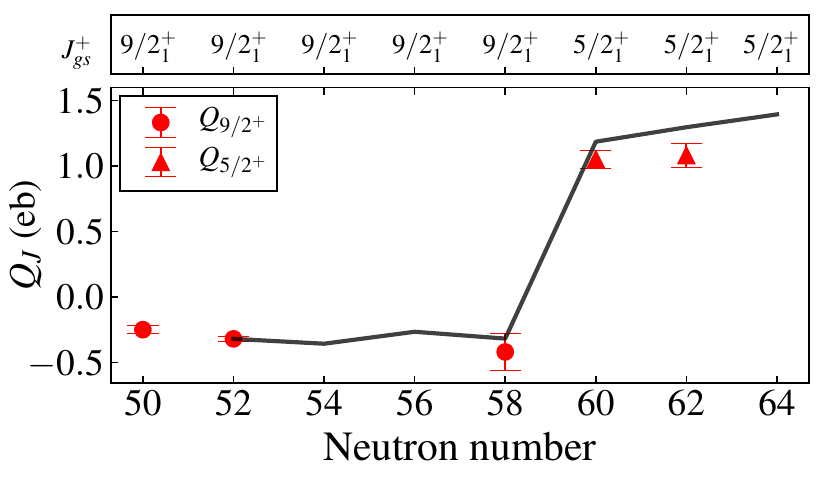}
\put (90,40) {(e)}
\end{overpic}
\begin{overpic}[width=0.49\linewidth]{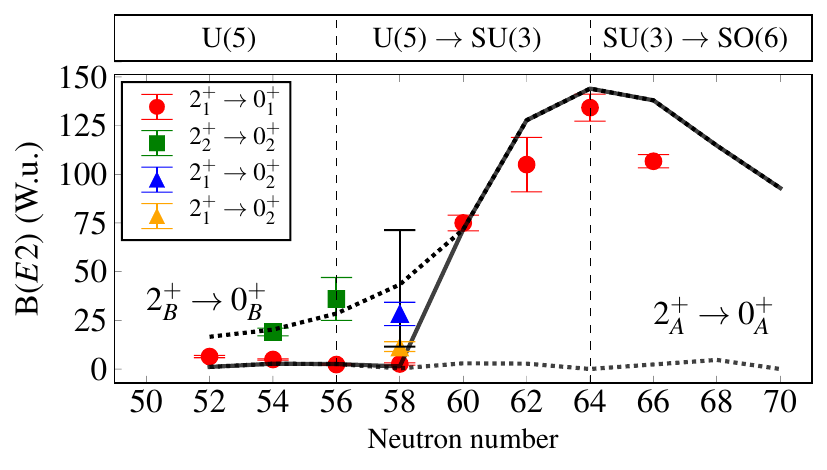}
\put (90,40) {(c)}
\end{overpic}
\begin{overpic}[width=0.49\linewidth]{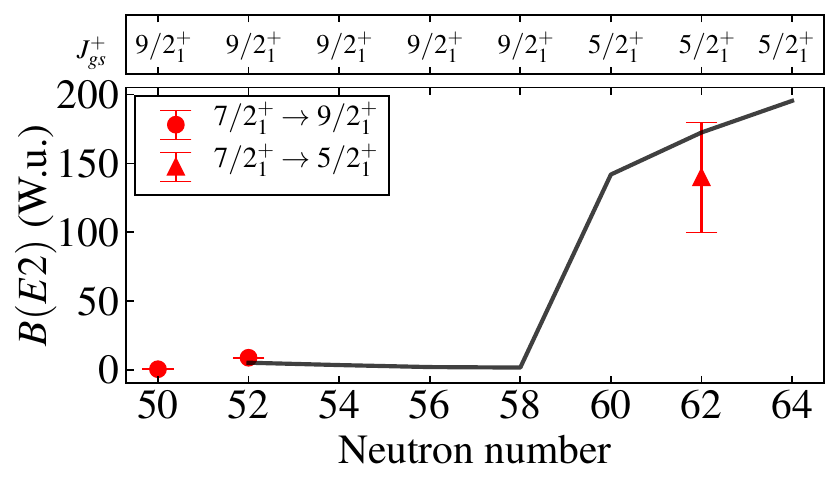}
\put (90,40) {(f)}
\end{overpic}
\caption{Evolution of spectral properties along the Zr
  and Nb chains.
  Symbols (solid lines) denote experimental data
  (calculated results).
  Left panels $^{92-110}$Zr isotopes.
  (a)~Percentage of the wave function within the
  intruder B-configuration for the ground ($0^+_{1}$)
  and excited ($2^+_{1}$) states.
  (b)~Normalized order parameters (see text for details).
  (c)~B(E2) values for $2^+\!\to\!0^+$ transitions
  in Weisskopf units (W.u.). Dotted 
lines denote calculated $E2$ transitions within a 
configuration. For the data,
see Fig.~17 of~\cite{Gavrielov2022}.
Right panels $^{93-105}$Nb isotopes.
(d)~Percentage of the intruder~(B) 
component [the $b^2$ probability in 
Eq.~(\ref{eq:norm_int})], in the ground state ($J^+_{gs}$) 
and the first-excited positive-parity state ($7/2^+_1$).
The values of $J^+_{gs}$ are indicated at the top.
(e)~Quadrupole moments of $J^+_{gs}$ in $eb$.
(f)~$B(E2; 7/2^+_1\!\to\! J^{+}_{gs})$ in W.u. 
For the data,
see Fig.~12 of~\cite{gavleviac22}.
}
\label{fig-order-p-be2}
\end{figure*}
\begin{figure*}
\centering
\includegraphics[width=1\linewidth]{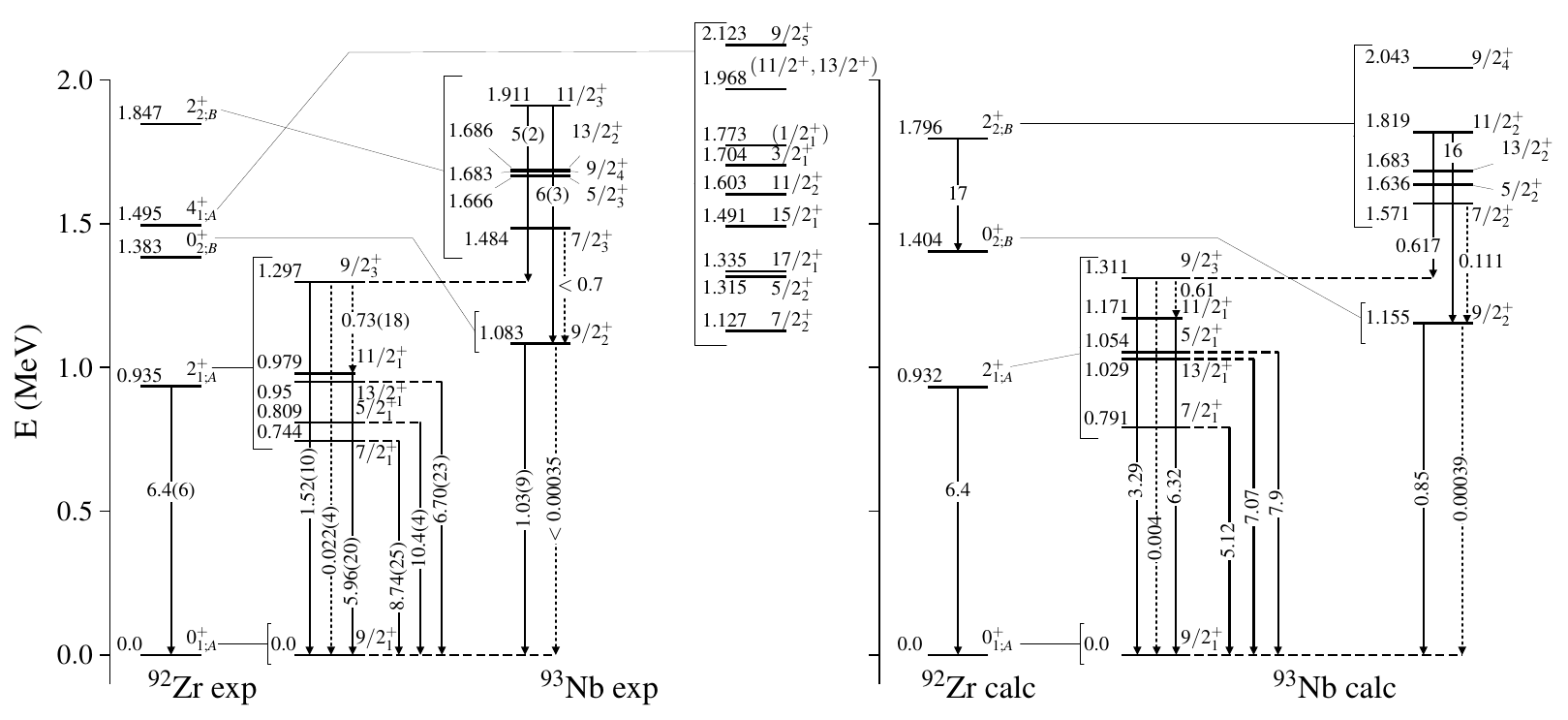}
\caption{Experimental (left) and calculated (right) energy 
levels in MeV, and $E2$ (solid arrows) and $M1$ (dashed 
arrows) transition rates in W.u., for $^{93}$Nb and 
$^{92}$Zr. Lines connect $L$-levels in $^{92}$Zr to sets of 
$J$-levels in $^{93}$Nb, indicating the weak coupling 
$(L\otimes \tfrac{9}{2})J$.
For the data, see Fig.~3 of~\cite{gavleviac22}.
Note that the observed $4^+_{\rm 1;A}$ 
state in $^{92}$Zr is outside the $N\!=\!1$ model 
space.\label{fig:93Nb-p}}
\end{figure*}

\section{IBFM with configuration mixing}
\label{sec-IBFMCM}
Odd-A nuclei are treated in the interacting boson-fermion
model (IBFM)~\cite{IBFMBook}, as a system of monopole ($s$) 
and quadrupole ($d$) bosons, representing valence nucleon 
pairs, and a single (unpaired) nucleon. The corresponding
Hamiltonian has the form
\ba
  \label{eq:ham}
\hat H = \hat H_{\rm b} + \hat H_{\rm f} + \hat V_{\rm bf} 
~.
\ea
For a single configuration, the boson part
involves an interacting boson model (IBM)~\cite{IBMBook}
Hamiltonian, taken to be
$ \hat H_{\rm b}(\epsilon_d,\kappa,\chi) =
  \epsilon_d\,
  \hat n_d + \kappa\,
  \hat Q_\chi \cdot \hat Q_\chi$,
  where $\hat{n}_d=\sum_{m}d^{\dag}_md_m$,
  $\hat Q_\chi =
d^\dag s+s^\dag \tilde d +\
\chi\, (d^\dag \tilde d)^{(2)}$ and
$\tilde{d}_m\!=\!(-1)^{m}d_{-m}$.
The control parameters $(\epsilon_d,\kappa,\chi)$
interpolate between the U(5) [spherical vibrator],
SU(3) [axial rotor] and SO(6) [$\gamma$-unstable rotor]
dynamical symmetry limits of the IBM.
The fermion part involves the single-particle term,
$\hat H_{\rm f} = \sum_j\epsilon_j\,\hat{n}_j$.
The boson-fermion part involves monopole, quadrupole
and exchange terms, 
$\hat V_{\rm bf} \!=\!
V_{\rm bf}^{\rm MON} + \hat{V}_{\rm bf}^{\rm QUAD}
+ \hat{V}_{\rm bf}^{\rm EXC}$.
For~a~single-j fermion,
$V_{\rm bf}^{\rm MON} \!=\! A\,\hat{n}_d\,\hat{n}_j$, 
$\hat{V}_{\rm bf}^{\rm QUAD} \!=\! 
\Gamma\,\hat{Q}_{\chi}\cdot
( a_{j}^{\dag }\, \tilde{a}_{j} )^{(2)}$,
$\hat{V}_{\rm bf}^{\rm EXC} \!=\! 
\Lambda
\sqrt{2j+1}:[ ( d^{\dag }\, \tilde{a}_{j})^{(j)}\times
  ( \tilde{d}\, a_{j}^{\dag })^{(j)}]^{(0)}:$ 
where $\tilde{a}_{j,m} \!=\! (-1)^{j+m}a_{j,-m}$
and  $:\,:$ denotes normal ordering.
The IBM and IBFM Hamiltonians,
$\hat H_{\rm b}(\epsilon_d,\kappa,\chi)$ and $\hat{H}$
of Eq.~(\ref{eq:ham}), have been used extensively
in the study of Type~I QPTs in even-even
nuclei~\cite{Cejnar2010,Casten2009,Iachello2011,
  Fortunato2021} and odd-even
nuclei~\cite{ScholtenBlasi1982,IBFMBook,Petrellis2011a,
  Petrellis2011b,Nomura2016a,Nomura2020,Boyukata2021}),
respectively.

For two configurations (A,B), the Hamiltonian of
the interacting boson-fermion model with configuration
mixing (IBFM-CM)
can be cast in matrix form~\cite{gavleviac22},
\ba
\hat H =
\begin{bmatrix}
  \hat H^{\rm A}_{\rm b}
+ \hat{H}_{\rm f} + \hat{V}_{\rm bf} &
\hat{W}_{\rm b} + \hat{W}_{\rm bf}\\
\hat{W}_{\rm b} + \hat{W}_{\rm bf}
& \hat H^{\rm B}_{\rm b}
+ \hat{H}_{\rm f} + \hat{V}_{\rm bf}
\end{bmatrix} ~.
\label{Hibfm-cm}
\ea
Here $\hat H^{\rm A}_{\rm b}
\!=\! \hat H_{\rm b}(\epsilon^{(A)}_d,\kappa^{(A)},\chi)$
represents the normal A configuration ($N$ boson space)
and $\hat H^{\rm B}_{\rm b}
\!=\! \hat H_{\rm b}(\epsilon^{(B)}_d,\kappa^{(B)},\chi)
+ \kappa^{\prime(B)} \hat L\cdot\hat L + \Delta_p$,
with an added rotational term,
represents the intruder 
B~configuration ($N\!+\!2$ boson space), corresponding to 
2p-2h excitations across the (sub-) shell closure.
For simplicity, $\hat{H}_{\rm f}$ and $\hat{V}_{\rm  bf}$
are taken to be the same in both configurations.
The boson- and boson-fermion mixing terms are
$\hat W_{\rm b} +\hat{W}_{\rm bf} \!=\!
[\omega + \sum_j\omega_j \hat n_j ]
[(d^{\dag}d^{\dag})^{(0)}
  \!+\! (s^{\dag})^2 + \text{H.c.}]$, where H.c. stands
for Hermitian conjugate. For
$\hat{H}_{\rm f} = \hat{V}_{\rm bf} = \hat{W}_{\rm bf}=0$,
the IBFM-CM Hamiltonian of Eq.~(\ref{Hibfm-cm}) reduces
to that of the IBM-CM model~\cite{Duval1981, Duval1982},
widely used in the study of configuration-mixed QPTs
and shape-coexistence in even-even
nuclei~\cite{Duval1981, Duval1982, 
  Sambataro1982, Ramos2014, Nomura2016c, Lev2018,
  Ramos2019, Gavrielov2019,Gavrielov2020,Gavrielov2022}.

The eigenstates
of $\hat{H}$~(\ref{Hibfm-cm}), $\ket{\Psi;J}$,
are linear combinations of wave functions $\Psi_{\rm A}$ 
and $\Psi_{\rm B}$, involving bosonic basis states in the 
two spaces $\ket{[N],\alpha,L}$ and $\ket{[N+2],\alpha,L}$.
Here $\alpha$ denote additional quantum~numbers of the
dynamical symmetry chain.
The boson ($L$) and fermion ($j$) angular momenta are 
coupled to $J$, $\ket{\Psi;J} \!=\! 
\sum_{\alpha,L,j}C^{(N,J)}_{\alpha,L,j}
  \ket{\Psi_{\rm A};[N],\alpha,(L\otimes j)J}
+ \sum_{\alpha,L_,j}C^{(N+2,J)}_{\alpha,L,j}
\ket{\Psi_{\rm B};[N+2],\alpha,(L\otimes j)J}$.
The probability of normal-intruder mixing is given by
\begin{equation}\label{eq:norm_int}
  a^2\!=\!\sum_{\alpha,L,j}|C^{(N,J)}_{\alpha,L,j}|^2,
  \;\;
  b^2\!=\!\sum_{\alpha,L,j}|C^{(N+2,J)}_{\alpha,L,j}|^2
\!=\!1-a^2.
\end{equation}
The $E2$ operator for a single-$j$ fermion has the form
$\hat{T}(E2) = e^{(\rm A)}\hat Q^{(N)}_{\chi}
+ e^{(\rm B)}\hat Q^{(N+2)}_{\chi}
+ e_f(a^{\dag}_j\,\tilde{a}_j)^{(2)}$,
where the superscript $(N)$ denotes a projection onto the 
$[N]$ boson space and
$(e^{(\rm A)},e^{(\rm B)},e_f)$ are effective charges.
The parameters of the IBFM-CM Hamiltonian and transition
operators are determined from a fit in the manner
described in~\cite{gavleviac22,Gav23}.

\section{QPTs in the Niobium chain}
\label{sec-Nb}
The $^{A}_{41}$Nb
isotopes with mass number 
\mbox{$A\!=\!\text{93--105}$} are described by coupling a 
proton to their respective $_{40}$Zr cores with
neutron number 52--64 (boson numbers $N$=1--7).
In the latter, the normal 
A~configuration corresponds to having no active protons 
above the $Z\!=\!40$ sub-shell gap, and the intruder 
B~configuration corresponds to two-proton excitation from 
below to above this gap, creating 2p-2h states.
In the present contribution,
we focus on the positive-parity 
states in the Nb isotopes~\cite{gavleviac22}
(both parity states are addressed in~\cite{Gav23}).
Such a case, reduces to a 
single-$j$ calculation, with the $\pi(1g_{9/2})$ orbit 
coupled to the boson core.
The IBFM-CM model space employed,
consists of a single-fermion plus
$[N]\oplus[N+2]$ boson spaces with
$N\!=\!1,2,\ldots,7$ for $^{93-105}$Nb.
The two configurations relevant for $^{99}$Nb
are shown schematically in Fig.~1.

Figures 2(b) and 2(c) show the experimental
and calculated levels of selected states
in the Nb isotopes
along with assignments to configurations based on 
Eq.~(\ref{eq:norm_int}). Empty (filled) symbols 
indicate a dominantly normal (intruder) state with small 
(large) $b^2$ probability. In the region between neutron 
number 50 and 56, there appear to be two sets of levels 
with weakly deformed structure, associated with 
configurations A and B. All levels decrease in energy 
for 52--54, away from closed shell, and rise again 
at 56 due to the $\nu(2d_{5/2})$ subshell closure. From 58, 
there is a pronounced drop in energy for the states of the 
B~configuration. At 60, the two configuration cross, 
indicating a Type~II QPT, and the ground state changes from 
$9/2^+_1$ to $5/2^+_1$, becoming the bandhead of a 
$K=5/2^+$ rotational band composed of  $5/2^+_1, 7/2^+_1, 
9/2^+_1, 11/2^+_1, 13/2^+_1$ states. The intruder 
B~configuration remains strongly deformed and the band 
structure persists beyond 60. As shown in Fig.~2(a),
a similar trend is encountered in the
even-even Zr isotopes with the same 
neutron numbers, where the spherical-to-deformed
Type~I QPT within the intruder configuration is
associated with U(5)-SU(3) transition and subsequently
SU(3)-SO(6)
crossover~\cite{Gavrielov2019,Gavrielov2020,Gavrielov2022}.

A possible change in the angular momentum of the ground 
state ($J^{+}_{gs}$) is a characteristic signature of 
Type~II QPTs in odd-mass, unlike even-even nuclei where the 
ground state remains $0^+$ after the crossing. It is an 
important measure for the quality of the calculations, 
since a mean-field approach, without configuration mixing, 
fails to reproduce the switch from $9/2^+_1$ to
$\!5/2^+_1$ in $J^{+}_{gs}$ for the Nb
isotopes~\cite{Guzman2011}. 
Fig.~3(d) shows the percentage of the wave function within 
the B~configuration for $J^+_{gs}$ and $7/2^+_1$, as a 
function of neutron number across the Nb chain. The rapid 
change in structure of $J^+_{gs}$ from the normal 
A~configuration in $^{93-99}$Nb (small $b^2$ probability)
to the intruder B~configuration in $^{101-105}$Nb (large 
$b^2$) is clearly evident, signaling a Type~II QPT. The 
configuration change appears sooner in the $7/2^+_1$ state, 
which changes to the B~configuration already in $^{99}$Nb.  
Outside a narrow region near neutron number 60, where the 
crossing occurs, the two configurations are weakly mixed 
and the states retain a high level of purity.
As shown in Fig.~3(a), a similar trend is
encountered for the $0^+_1$ and $2^+_1$  
states in the respective $_{40}$Zr cores.

Further insight into the nature of the QPTs is gained by
considering the behaviour of the order parameters and
related observables. The quadrupole moment of $J^{+}_{gs}$
and $B(E2; 7/2^+_1\!\to\! J^{+}_{gs})$
in Nb isotopes are shown
in Fig.~3(e) 
and Fig.~3(f), respectively.
These observables are related to the deformation, the order 
parameter of the QPT. Although the data is incomplete, one 
can still observe small (large) values of these observables 
below (above) neutron number 60, indicating an increase in 
deformation. The calculation reproduces well this trend and 
attributes it to a Type~II QPT involving a jump between 
neutron number 58 and 60, from a weakly-deformed 
A~configuration, to a strongly-deformed B~configuration. 
This behavior is correlated with 
a similar jump seen for the 
B(E2)'s of $2^+\!\to\!0^+$ transitions in the even-even
Zr cores, Fig.~3(c), and with the calculated order
parameters, Fig.~3(b). The latter are
the expectation value of
$\hat{n}_d$ in the $0^{+}_1$ ground state wave function,
$\braket{\hat{n}_d}_{0^{+}_1}$,
and in its $\Psi_A$ and $\Psi_B$ components,
$\braket{\hat{n}_d}_A$, and $\braket{\hat{n}_d}_B$.
Their evolution along the Zr chain reveals that
configuration~A remains spherical, while configuration~B
undergoes a Type~I QPT involving a gradual
spherical-to-deformed [U(5)-SU(3)-SO(6)] shape-phase
transition.

Additional evidence for a \mbox{Type~I} QPT,
involving shape changes  
within the intruder B~configuration,
is obtained by examining the
individual structure of Nb isotopes at the end-points of 
the region considered. Fig.~\ref{fig:93Nb-p} displays the 
experimental and calculated levels in $^{93}$Nb along with 
$E2$ and $M1$ transitions among them. The corresponding 
spectra of $^{92}$Zr, the even-even core, are also shown  
with an assignment of each level $L$ to the normal A or 
intruder B configurations, based on the analysis 
in~\cite{Gavrielov2022}, which also showed that the two 
configurations in $^{92}$Zr are spherical or 
weakly-deformed. It has long been known~\cite{Heerden1973}, 
that low-lying states of the A~configuration in $^{93}$Nb, 
can be interpreted in a weak coupling scheme, where the 
single-proton $\pi(1g_{9/2})$ state is coupled to 
spherical-vibrator states of the core. Specifically, for 
the $0^+_{1;A}$ ground state of $^{92}$Zr, this coupling 
yields the ground state $9/2^+_1$ of $^{93}$Nb. For 
$2^+_{1;A}$, it yields a quintuplet of states, $5/2^+_1, 
7/2^+_1, 9/2^+_3, 11/2^+_1,13/2^+_1$, whose ``center of  
gravity'' (CoG)~\cite{Lawson1957}, is 0.976~MeV, in 
agreement with the observed energy 0.935~MeV of $2^+_1$
in $^{92}$Zr. The $E2$ transitions from the quintuplet  
states to the ground state are comparable in magnitude
to the $2^+_{1;A}\!\to\!0^+_{1;A}$ transition in $^{92}$Zr, 
except for $9/2^+_3$, whose decay is weaker.
A nonet of states built on  
$4^+_{1;A}$ can also be identified in the empirical 
spectrum of  $^{93}$Nb, with CoG of 1.591~MeV, close to 
1.495~MeV of $4^+_{1;A}$.
\begin{figure}[b]
\centering
\includegraphics[width=\linewidth]{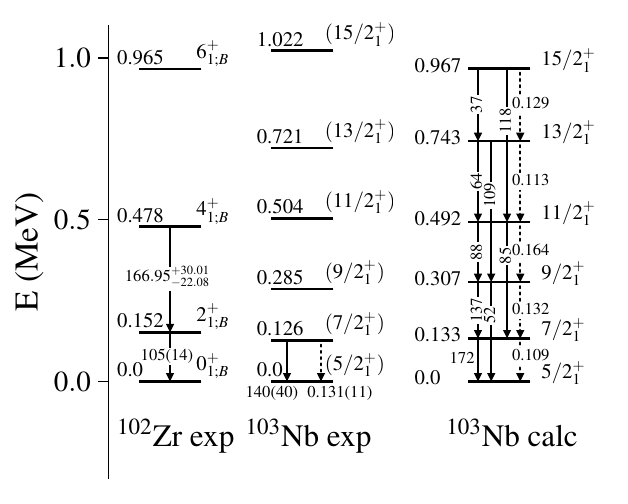}
\caption{Experimental and calculated energy 
levels in MeV, and $E2$ (solid arrows) and $M1$ (dashed 
arrows) transition rates in W.u., for $^{103}$Nb and
$^{102}$Zr. For the data,
see Fig.~4 of~\cite{gavleviac22}.
\label{fig:103Nb-p}}
\end{figure}

Particularly relevant to the present discussion is the
fact that the weak-coupling scenario is also valid for
non-yrast states of the intruder B configuration in
$^{93}$Nb. As shown in Fig.~\ref{fig:93Nb-p},
the coupling of $\pi(1g_{9/2})$ to 
the $0^+_{2;B}$ state in $^{92}$Zr, yields the excited 
$9/2^+_2$ state in $^{93}$Nb. For $2^+_{2;B}$~it yields the 
quintuplet, $5/2^+_3, 7/2^+_3, 9/2^+_4, 11/2^+_3, 
13/2^+_2$, whose CoG is 1.705~MeV, a bit lower than 
1.847~MeV of $2^+_{2;B}$. The observed $E2$ rates 
$1.03(9)$~W.u for \mbox{$9/2^+_2\!\to\!9/2^+_1$}, is
close to the calculated value 0.85~W.u., but is smaller
than the observed value $1.52(10)$~W.u for 
\mbox{$9/2^+_3\!\to\!9/2^+_1$}, suggesting that $9/2^+_2$ is 
associated with the B~configuration.

For $^{103}$Nb, the yrast states shown in
Fig.~\ref{fig:103Nb-p}
are arranged in a $K=5/2^+$ rotational band, with an 
established~\cite{Hotchkis1991} Nilsson model assignment 
$5/2^+[422]$. The band members can be interpreted in the 
strong coupling scheme, where a particle is coupled to an 
axially-deformed core. The indicated states are obtained by 
coupling the $\pi(1g_{9/2})$ state to the ground band 
($L=0^+_{1;B},2^+_{1;B},4^+_{1;B},
6^+_{1;B},\ldots$) of $^{102}$Zr, also shown in
Fig.~\ref{fig:103Nb-p},
which is  
associated with the intruder B~configuration. The 
calculations reproduce well the observed particle-rotor 
$J(J+1)$ splitting, as well as, the $E2$ and $M1$ 
transitions within the band. Altogether, we see an 
evolution of structure from weak-coupling of a spherical 
shape in $^{93}$Nb, to strong-coupling of a deformed shape 
in $^{103}$Nb. Such shape-changes within the 
B~configuration (Type~I QPT), superimposed on abrupt 
configuration crossing (Type-II QPT), are the key defining 
feature of intertwined QPTs. Interestingly, this  
intricate scenario, originally observed in the 
even-even Zr
isotopes~\cite{Gavrielov2019,Gavrielov2020,Gavrielov2022},
persists in the adjacent odd-even Nb~isotopes.

\section{Conclusions}
\label{sec-concl}
An application of the recently introduced
IBFM-CM framework~\cite{gavleviac22,Gav23} to
Nb isotopes, disclosed a Type-II QPT
(abrupt crossing of normal and intruder configurations)
accompanied by a Type~I QPT
(gradual shape-evolution
and transition from weak to strong coupling within
the intruder configuration), thus 
demonstrating, for the first time, intertwined QPTs
in odd-mass nuclei.
The observed concurrent types of QPTs
in odd-even Nb isotopes echo the
intertwined QPTs previously found in the adjacent  
even-even Zr
isotopes~\cite{Gavrielov2019,Gavrielov2020,Gavrielov2022}. 
The results obtained motivate further experiments of 
non-yrast spectroscopy in
such nuclei and their quantitative, yet
physically transparent, description
in the IBFM-CM framework.

This research was supported
by the US-Israel Binational Science Foundation
Grant No.~2016032.

\end{document}